# Studies of YBCO Strip Lines under Voltage Pulses: Optimisation of the Design of Fault Current Limiters

M. Decroux, L. Antognazza, S. Reymond, W. Paul, M. Chen and Ø. Fischer

*Abstract*— We present experimental results on the behaviour of a superconducting YBCO/Au meander of length L submitted to short circuit tests with constant voltage pulses. The meander, at the beginning of the short-circuit, is divided in two regions; one, with a length $L_1$ proportional to the applied voltage, which first switches into a highly dissipative state (HDS) while the rest remains superconducting. Then the rest of the meander will progressively switch into the normal state due to the propagation of this HDS (few m/s) from both ends. The part $L_1$ has to initially support a power density proportional to $r \cdot J_p^2$ ($r$ is the resistivity of the bilayer and $J_p$ the peak current density). To avoid local excessive dissipation of power and over heating on one part of the wafer in the initial period, we have developed a novel design in order to distribute the dissipating section of the meander into many separated small dissipative zones. Furthermore the apparent propagation velocity of these dissipative zones is increased by the number of propagation fronts. We will show results obtained on 3kW ( 300V, 10A) FCL on a 2" wafer which confirm the benefits of this new design.

## I. Introduction

Fault current limiter (FCL) are still one of the most promising application of high $T_c$ superconductors since it has no counterpart in the actual non superconducting technology. Even if there are two kinds of FCL, namely the inductive and the resistive types, the latter one seems to take the lead as it is under development by the majors electrical companies in Europe. ABB has already built a 6.4MVA FCL, the highest rated power for HTS, made of BiSCCO ceramic meanders [1] and Schneider Electric is testing a 1MVA demonstrator with pellets of melt textured YBCO [2]. On the other hand, Siemens has recently produced a 1.25MVA FCL based on YBCO thin films [3]. All the thin films based FCL have a similar behavior; the switching time is very low, of the order of few microseconds, and the current peak during a short circuit reaches 3-4 times the critical current. However it has been observed that only a part of the FCL is actually limiting the current and also is supporting all the dissipated power [4], implying that there is still a room for the optimization of the FCL's performance. To improve these performances there is a need of understanding the quench properties of superconducting line. This can be done by carefully studying the behavior of YBCO line under high constant current pulses and voltage pulses [5]. In this paper we describe the result of these studies and, based on it, we propose a new design of FCL. We also present results obtained on a test wafer and on a 3KW FCL on 2" wafer.

## II. Experimental

The measurements are performed on Au\YBCO\CeO$_2$ heterostructures grown epitaxially onto 2" sapphire substrates [6]. The thickness of Au and YBCO layers are respectively 50-100 nm and 300nm. All the structures were processed by conventional wet etching. The critical temperature $T_c$ is 88K and the resistivity of the Au\YBCO bi-layer is $r$ =5-10 µΩcm at 90K. The standard critical current density $J_c$, measured both by an inductive method and by the I-V curves obtained with current pulses, is around 3 MA/cm$^2$ at 77K in self magnetic field, with a spatial homogeneity of ±10%.

## III. Results

### A. Current pulses

Our investigations on the behavior of superconducting line at high current densities have already been published elsewhere [5]. Briefly, when a constant current pulse is applied to the YBCO line we always observe that the line stays in a superconducting state for a certain delay time, before it quickly switches into a highly dissipative state. We have shown that this transition is not due to a thermal runaway of the line, but it is driven by the applied current, leading to a new kind of critical current [5,7]. It is still not clear if this HDS, when it occurs, is a flux flow state or the normal state. However, once the HDS is set, the temperature quickly increases leading to the normal state. The delay time decreases rapidly with an

Manuscript received August 6, 2002. This work was supported by CTI under contract N°3625.1

M. Decroux, L. Antognazza, S. Reymond, and Ø. Fischer are with the University of Geneva, DPMC, 24 quai Ernest Ansermet, 1211 Geneva 4, Switzerland (corresponding author: L. Antognazza phone: 4122-702-6305 fax: 4122-702-6869; e-mail: louis.antognazza@physics.unige.ch).

W.Paul and M. Chen are with ABB Corporate Research, Dättwil, Switzerland.

increasing current and a quasi instantaneous (<μs) transition takes place for $J$ of the order of $3·J_c$, which is also the peak current observed in FCL during a short circuit. Once part of the line has switched, the HDS starts to propagate. By recording the resistance of different adjacent sections of the line we were able to observe this propagation and to measure its current density dependence [5]. The propagation velocities increase with the current density from 20 m/s for $J\sim 1.5·J_c$ to 150 m/s for $J\sim 2.6·J_c$, and diverge as J approaches $3·J_c$. This current dependence can be well simulated by a model which takes into account the existence of this new critical current [8].

*B. Voltage pulses*

When a voltage pulse is applied to a superconducting line we initially observe the usual FCL behavior; i.e. a transient regime where the current is limited after few microseconds and where the peak current density reaches values of $J_p \approx 3·J_c$. The line has then to support an initial surface peak power density of $P_{peak} = r·J_p^2·d \approx 12\text{-}25 kW/cm^2$ depending on the gold layer thickness. After this transient regime we notice that the resistance of the line, e.g. after 25 μs, varies linearly with the applied voltage, indicating that the initial length of the HDS, which is directly related to this resistance, is proportional to this voltage [5]. This implies that at these times (typically around 25μs), the current decreases to a value $J_t \sim 1.5·J_c$ which is independent of the applied voltage. From this characteristic we can define a critical electric field $E_c$ of the line; $E_c = r·J_t \sim 1.5·r·J_c$, where ρ is the resistivity above $T_c$, as after 25μs the HDS is in the normal state. As we will discuss hereafter, $E_c$ is one of the key parameters for the design of a FCL: it depends essentially on the thickness of the gold layer and in our geometry $E_c \sim 25\text{-}50 V/cm$. However $E_c$ can not be chosen to high, since the dissipated power density after the transient regime is directly related to it ($P_t = r·J_t^2·d \sim 1.5·E_c·J_c·d$) and we have shown that values higher than $22 kW/cm^2$ can result in damage for the FCL [5].

*C. Design of the meander*

By taking into account these experimental facts, we can now sketch the behavior of a FCL (meander of length $L_{tot}$) during a short circuit. For a given short circuit voltage $U$, the current peak reaches $3·I_c$ and the length of the initial dissipative region is $L_1 = U/E_c$. The meander is then divided in two parts, as represented in Fig.1a: while the length $L_1$ is in the dissipative state the second part $L_2$ is still in the superconducting state.

After 25μs the current in the line is around $1.5·I_c$ and at this moment the HDS propagates with a velocity of ~ 20m/s. This velocity will then quickly decrease since the propagation of the HDS leads to a further decrease of the current. In our geometry the current goes down, in less than 100μs, to $I_c$ at which the propagation is only of the order of 1-2 m/s. Due to this slow propagation, the second part of the meander will transit into the normal with a rate of only 2-4 m/s (since there are two fronts of propagation).

It is obvious that a design where this length $L_1$, obtained for the maximum nominal voltage $U_{max}$, is significantly lower than the total length of the meander is not optimal in that only a part of the FCL is participating in the initial limitation of the current. Furthermore the dissipated power will be located in one part of the meander which could limit the maximal power sustained by one wafer. On the other hand if the length $L_1 > L_{tot}$, both the peak current and the current after the transient regime will be higher than in the previous case resulting in a possible damage for the meander. These points underline the importance for FCL of the length $L_1$ and therefore of $E_c$.

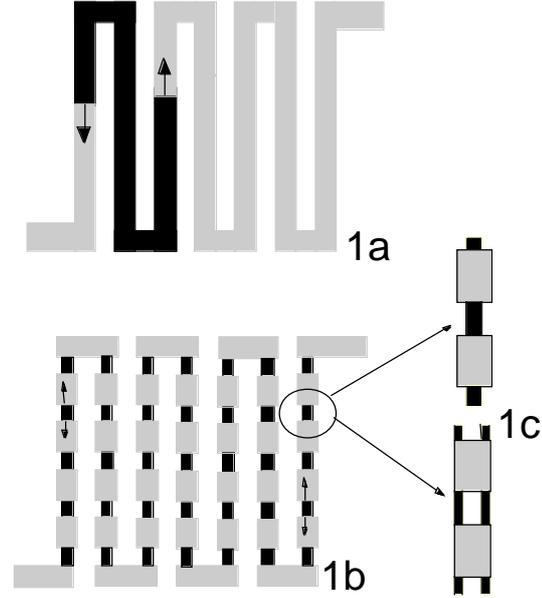

Fig. 1a: In a standard design the meander initially switches over a length (in black) $L_1 = U/E_c$. 1b: the new design distributes this dissipative zone into many separated small dissipative zones (constrictions, in black). 1c: The constrictions can be split into two parts in order to improve its thermal stability. The arrows in the meander are indicating the propagation of the highly dissipative state.

To avoid the problems related to the localization of the HDS in one part of the meander, we propose a new design in which the initial length $L_1$ of the HDS is split and uniformly distributed along the meander, as represented in Fig. 1b. This is done by locally decreasing the width of the line (and therefore the critical current), i.e. by including constrictions, of width $w_c$, along the meander. These constrictions can also be split in two parts of width $w_c/2$, as in Fig. 1c; this increases the heated volume in the wafer and then improves their thermal stability. The meander should be design in such a way that only the constricted parts are switching during the first microseconds of a short circuit and that, for $U_{max}$, the total length $L_c$ of the constrictions switches, i.e. $L_c = U_{max}/E_c$.

There are two main benefits of this new design; first the initial dissipated power will be homogeneously distributed along the meander. In the case of FCL based on several wafers in series $L_1$ is distributed over all the wafers, avoiding that only one of the wafer has to support all the dissipated power. The second advantage is resulting from the increase of the number of propagation front; indeed the HDS is propagating from each constriction into the two adjacent connecting paths, implying that the total effective propagation velocity can be increased by a factor n, where n is the number of constrictions. The increase of the effective propagation velocity speeds up the increase of the resistance line which, in turn, speeds up the decrease of the current and of the dissipated power in the meander.

### D. Test wafer

In order to observe the effect of this new design on the performances of a FCL, we first produced a test wafer (100V/5.4A), presented in Fig.2. Voltage contacts are located in the middle of each connecting path, in order to have the possibility to measure the behavior of all the constricted parts at the same time.

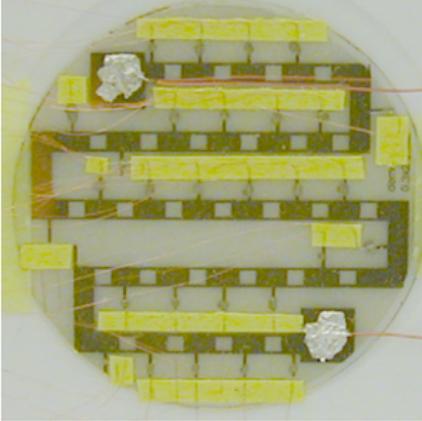

Fig 2: View of the 2" test wafer with the new design of the meander.

We already mentioned that only the constrictions should switch at the beginning of the short circuit. Since there is an inhomogeneity of the critical current density over the wafer this implies that the maximum critical current of the constrictions has to be lower than the minimum critical current of the connecting path. In other words the ratio between the width of the constrictions and the width of the connecting paths has to be higher than the ratio of the maximum and the minimum critical current density over the wafer.

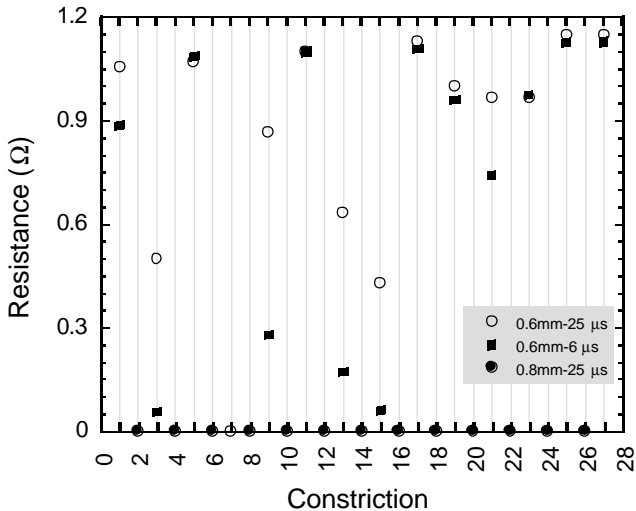

Fig.3: Resistance of the constrictions, after 6 and 25μs, for a voltage pulse of 100V.

To have an idea of what should be the difference in width of the constricted and the connecting paths, we designed this meander with 14 constrictions having a width of 2x0.3mm=0.6mm ($I_c$=5.4A) and 13 a width of 2x0.4mm=0.8mm ($I_c$=7.2A). All the constricted parts are 2mm long, and the connecting path are 4mm long by 2mm width.

Fig 3 shows the resistance of each constriction, 6 and 25 μs after the beginning of a 100V voltage pulse. We can notice that only the constrictions with the smaller width (odd number) are in the dissipative state (except n°7) after 25μs, indicating that a 33% difference in width is enough to discriminate between the two types of constrictions. This implies that, for further design of FCL, we can use this difference in width between the constrictions and the connecting path.

Fig. 4 shows the behavior of all the 0.6mm width constrictions for a pulse of 100V and 20 ms.

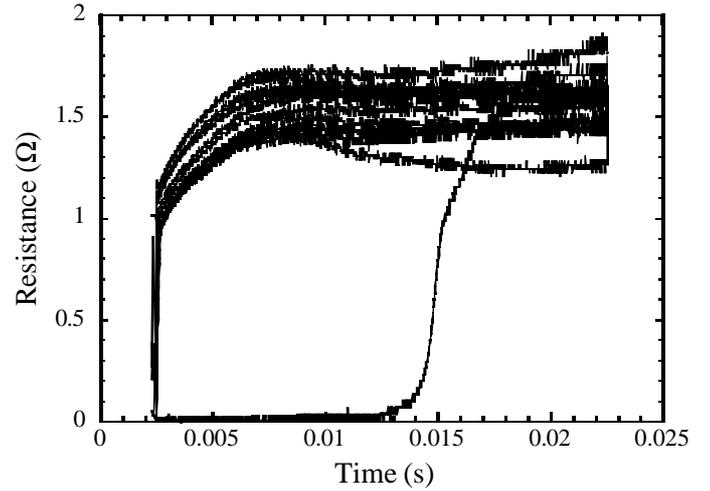

Fig 4: Resistance of all the 0.6mm width constrictions for a voltage pulse of 100V and 20ms.

All these constrictions (except n°7) have switched in few μs showing that the dissipated power is well distributed over all the meander. It is worth noting the good homogeneity of the resistances after 20ms. The 0.8mm width constrictions start their transition ~4 ms after the beginning of the pulse. This delay can be explained by the slow propagation, of the order of 1 m/s (since the current is below $I_c$), of the HDS through the connecting path (4mm long).

### E. 3 kW FCL

Since the test wafer gave good results, we have then tested a 3kW (300V/10A) and a 5kW (300V/16A) FCL on 2 " wafer. We present here the results obtained on the 3kW FCL, the meander is shown in the inset of Fig. 5. The constrictions, with a length of 2mm and a width of 1mm ($I_c$=10A) are homogeneously distributed along the 8 lines of the meander. By choosing a critical field of $E_c$= 25V/cm, the total length of the constricted parts is fixed to $L_c$= U/$E_c$=12cm. The connecting paths are 2mm x 4mm long.

When a 300V voltage pulse is applied to the meander, the peak current reaches the usual value of 3 $I_c$ and then goes quickly down to $I_c$ in about 40μs. 6 of the 8 lines have initially switched into the HDS, the two other lines are still in the superconducting state, indicating that certainly the value of the critical field $E_c$ is higher than the one expected. Since $E_c$ depends essentially of the gold layer, this shows the importance of having a good control of the thickness and of the quality of this layer. It is interesting to notice that for the same meander but without constrictions, the initial length of the HDS (12 cm) should cover, at most, 3 lines compared to the 6 lines observed with this new design.

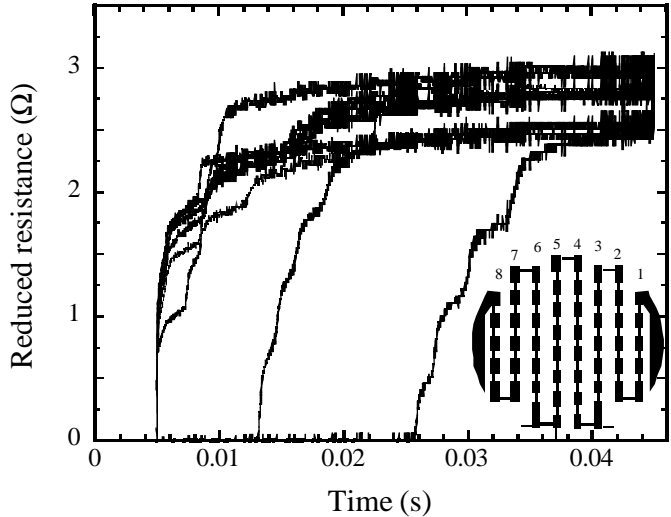

Fig. 5: Resistance of the 8 lines of the meander during a voltage pulse of 300V and 40ms. This resistance is divided by the number of constriction in each line. The inset shows the design of the meander.

Even with this problem the meander sustained many 300V pulses, without any damage, for time as long as 100ms. Fig. 5 shows the results for a 300V/40ms pulse. Since each line has a different length and a different number of constrictions, we divided the resistance of the line by the number of constriction in this line. The line 1 and 2 are starting to switch after 8ms and 20ms respectively; at these times the current is well below $I_c$ which means that this transition is due to the slow (1m/s) propagation of both the HDS and the heat in the substrate. This slow switching however allows to see more clearly the staircase like structures of the transition, also observed in all the other lines. These structures are resulting from the successive transition of the constrictions and of the connecting path. The transition of the connecting path is slower than in the constrictions since the current density is smaller.

After 40ms one observes that all the lines have a similar reduced resistance, suggesting a good homogeneity of the dissipated power in the meander. This homogeneity is confirmed in Fig. 6 which shows the averaged surface power density for all the lines at different times. After 30μs the power is sustained only by the constriction and it varies between 2.8 to 4.5 kW/cm$^2$ from the initial peak power of $P_{peak}$ = 12kW/cm$^2$. As the time increases the power distribution becomes more homogeneous and for t=40ms it is ranging from only 65 to 100W/cm$^2$.

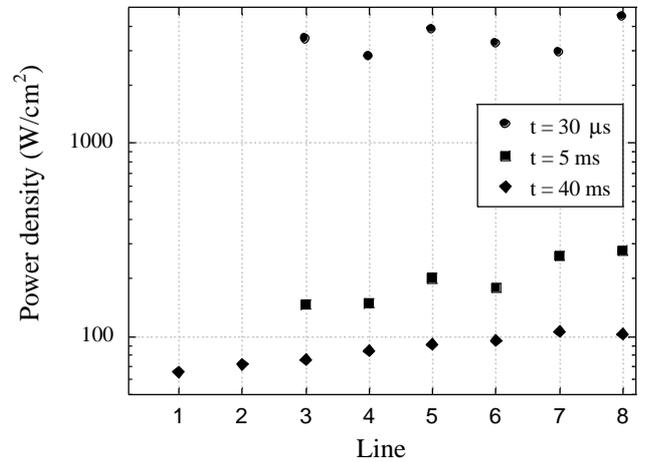

Fig. 6: Averaged surface power density, for all the lines of the meander at different times, for voltage pulses of 300V.

## IV. CONCLUSION

Based on the observed behavior of a YBCO line submitted to voltage and current pulses, we proposed and tested a new design for FCL. The main idea is to split the initial dissipative length into many small dissipative regions, which avoids the localization of the sustained power in one region of the wafer. This is realized by including constrictions, i.e. local decrease of the critical current, along the meander. We have tested this design on a test wafer and on a 3kW (300V/10A) FCL. The results show that the dissipated power is homogeneously distributed over the meander. The design of the wafer can be further improved by carefully choosing, for given nominal voltage and current, the value of $E_c$ and the optimal sizes of the constricted and connecting regions.